\documentclass[11pt,cite,a4paper]{article}
\pdfoutput=1
\usepackage{amsmath}
\usepackage{amssymb}
\usepackage{a4wide}
\usepackage{amssymb}
\usepackage{graphicx}
\usepackage{mathrsfs}

\newcommand{\be}{\begin {equation}}
\newcommand{\ee}{\end {equation}}
\newcommand{\ba}{\begin{eqnarray}}
\newcommand{\ea}{\end{eqnarray}}

\usepackage{titlesec}
\usepackage{titletoc}
\titleformat{\chapter}[display]
 {\normalfont\Large\bfseries}{\thechapter}{11pt}{\Large}\titleformat{\section}{\normalfont\large\bfseries}{\thesection}{11pt}{\large}
\usepackage[titletoc]{appendix}
\usepackage[colorlinks,linkcolor=red,anchorcolor=blue,citecolor=green]{hyperref}

\titlespacing*{\chapter}{0pt}{0pt}{15pt} 
\titlespacing*{\section}{0pt}{3.5explus1exminus.2ex}{2.3explus.2ex}
\usepackage{nameref}

\titleformat{\chapter}[display]
{\normalfont\Large\bfseries}{Appendix~\Alph{chapter}}{11pt}{\Large}

\numberwithin{equation}{section}

\begin{document}
 \vspace{2truecm}

\centerline{\Large \bf Entanglement Entropy Evolution under Double-trace Deformation}

\vspace{2.3truecm}

\centerline{
    {Yushu Song}
    }
\vspace{.8cm} \centerline{{\it College of Physical Science and Technology, Hebei University, Baoding 071002, China}}

\vspace*{2.0ex} \centerline{E-mail: {\tt yssong@hbu.edu.cn }}

\vspace{2.5truecm}

\centerline{\bf ABSTRACT}
\vspace{.5truecm}
In this paper, we study the bulk entanglement entropy evolution in conical BTZ black bole background using the heat kernel method. This is motivated by exploring the new examples where quantum corrections of entanglement entropy give the leading contribution. We find that in the large black hole limit the bulk entanglement entropy decreases under the double-trace deformation which is consistent with the holographic $c$ theorem and in the small black hole limit the bulk entanglement entropy increases under the double-trace deformation.

 \noindent

\newpage

\section{Introduction}
Entanglement is a mysterious property of quantum system that distinguishes them from their classical counter parts. In quantum mechanics entanglement leads decoherence of the superposition of the state which is central to the quantum measurement problem. Entanglement entropy is one of the widely used tool for describing the entanglement structure of many-body systems.

Generally entanglement entropy has a holographic interpretation. It was conjectured by Ryu and Takayanagi \cite{Ryu:2006bv} that the entanglement of a region $A$ in a CFT is given by the area of the minimal surface in Planck unit whose asymptotic boundary is the boundary of $A$. This proposal was proved for general case in \cite{Lewkowycz:2013nqa}. The explicit calculation of the quantum corrections to the Ryu-Takayanagi proposal were analyzed in \cite{Faulkner:2013ana} \cite{Dong:2012se}.

Motivated by the work of Miyagawa, Shiba and Takayanagi \cite{Miyagawa:2015sql} who studied the holographic entanglement entropy evolution in the AdS space, We will generalize the bulk entanglement evolution to black hole background. The main purpose of this paper is to study a example where quantum corrections of entanglement entropy give the leading contribution. We will study the evolution of bulk entanglement entropy in the conical BTZ black hole background under boundary double-trace deformation. In our holographic set-up, we simply choose the subsystem $A_b$ in the bulk to be the region inside horizon. In the boundary, from thermal field double point of view, the corresponding region $A$ is half of the whole boundary. In the bulk we consider the bulk entanglement entropy evolution for region $A_b$ whose classical value is the Bekenstein-Hawking entropy. For the case with $U(1)$ symmetry in the bulk, there are two methods to calculate the quantum corrections to entanglement entropy, the Euclidean path integral method \cite{Gibbons:1976ue} and off-shell method \cite{Frolov:1995xe} \cite{Solodukhin:2011gn}. In the Euclidean path integral approach, as we change the period of Euclidean time $\tau$, we always consider the smooth solution of the quantum corrected Einstein equations. In this approach, the Euclidean path integral determines the full quantum corrections which include the bulk entanglement entropy, area shift, Wald entropy and counterterms. In the off-shell approach, we treat the period of Euclidean time $2\pi \beta$ and the temperature $T$ as independent variables, and so varying $\beta$ is equivalent to introducing a conical singularity at horizon with deficit angle
$\delta = 2 \pi(1-\alpha)$ and $\alpha=\beta/\beta_H $ where $2\pi \beta_H$ is the inverse of Hawking temperature. In the off-shell approach, we can think of the theory in bulk as an effective field theory living in a fixed background, and so the off-shell approach essentially leads to the bulk entanglement entropy. In this paper we will calculate the bulk entanglement entropy using the off-shell approach and find out the evolution of the bulk entanglement entropy under the boundary renormalization group flow triggered by the double-trace deformation.

 This paper is organized as follows. In section 2, we introduce the bulk geometry in our set-up and the holographic description of the double-trace deformation. In section 3, we compute the bulk entanglement entropy in conical BTZ black bole background using the heat kernel method. We also derive the entropy difference under boundary renormalization group flow. In section 4, we present the conclusions. In the appendix, we review quantum correction to the entropy of BTZ black hole using the heat kernel method.

\section{Scalar in BTZ Background and Double-Trace Deformation}

We start with the Euclidean metric of three dimensional black hole in AdS space \cite{Banados:1992wn} which can be written as

 \ba
 ds^2= \frac{r^2-r_+^2}{l^2}d\tau^2+\frac{l^2}{r^2-r_+^2}dr^2+r^2d\phi^2,\nonumber
 \ea
 \ba \label{btz}
 \phi \sim \phi+2\pi,~~\tau\sim \tau+ 2\pi\beta_H ,~~\beta_H=\frac{ l^2}{r_+}
 \ea
 where $r_+\equiv l\sqrt{M}$ is the horizon of BTZ black hole which is related to the mass $M$ of black hole, $l$ is the radius of curvature of AdS$_3$ and
 $2\pi\beta_H$ is the inverse of Hawking temperature. The BTZ metric can be obtained from global AdS$_3$ by identifications

 \ba
 ds^2= l^2(d\rho^2+ \cosh^2\rho d\psi^2+ \sinh^2\rho d\theta^2), \nonumber
 \ea
 \ba \label{ads3}
\theta \sim \theta+2\pi l,~~\psi \sim \psi+ \frac{2\pi r_+}{l}\sim  \psi+ \frac{2\pi l}{\beta_H}
 \ea

Consider a massive scalar field $\varphi(r,x)$ in BTZ background which is dual to a scalar operator $O(x)$ in the boundary theory. Solving the equation of motion
for scalar field near the boundary $r\rightarrow \infty$, we obtain the asymptotical behavior of scalar field \cite{Klebanov:1999tb},
 \ba
& \varphi(r,x)=\alpha(x)r^{-\Delta_-}+\beta(x)r^{-\Delta_+},&
 \ea
where $\Delta_{\pm}=1\pm \sqrt{1+m^2l^2} $ is the conformal dimension of scalar operator $O(x)$. The standard boundary condition corresponds to $\alpha(x)=0$ and the alternate boundary condition corresponds to $\beta(x)=0$. For standard/alternate boundary condition, the dual scalar operator $O(x)$ has the conformal dimension $\Delta_+$/$\Delta_-$. Note that if we introduce $\nu=\sqrt{1+m^2l^2}$, shift of the boundary conditions corresponds to $\nu \rightarrow -\nu$. For a free scalar field of mass $m$ in AdS space, there are two possible ways of quantization if
 \ba
-1< m^2l^2<0 ,
 \ea
The unitary bound on the dimension of the scalar operator in $d$ dimensional boundary CFT requires $ 2\Delta > d-2$. In the following discussion, we will assume the unitary bound is always satisfied, that is $\Delta>0$. Suppose we start with CFT with the alternate boundary condition, i.e. CFT$_{A}$\footnote{This is defined as Neumann boundary condition CFT$_{N}$ in \cite{Miyagawa:2015sql} .}. The double-trace deformation of the CFT is given by
 \ba
\delta S=\lambda \int d^2 x O^2(x)
 \ea
For CFT with the alternate boundary condition the coupling constant $\lambda$ has dimension $2\nu$ which means the deformation is relevant. Under double-trace deformation, there is a renormalization group flow starting from a UV fixed point where $\lambda=0$. The end point of this flow is an IR fixed point where the boundary condition becomes standard boundary condition, i.e. CFT$_{S}$. In CFT$_{S}$, the dimension of $O^2(x)$ is $2\Delta_+$ and the deformation becomes irrelevant.

\section{Entanglement Entropy Evolution in BTZ Background}

\subsection{Heat Kernel Method}

The heat kernel \cite{Vassilevich:2003xt} is the fundamental solution to the heat equation on a specified domain with appropriate boundary conditions.
It is also one of the main tools in the study of the spectrum of Laplace operator. In general, the heat kernel exists on any Riemannian manifold with boundary provided the boundary is sufficient regular. The heat kernel can be formally defined as
\ba
K(s;x,y;D)=  \left \langle x | \exp(-sD)  | y \right \rangle
\ea
where $D\equiv-\nabla^2+m^2$. This expression means that $K$ should satisfy the heat equation with delta function initial condition
\ba
(\frac{\partial}{\partial s} + D)K(s;x,y;D)=0; \nonumber
\ea
\ba
K(0;x,y;D)=\delta(x,y)
\ea
The propagator $D^{-1}(x,y)$ can be defined through heat kernel by the integral representation if the heat kernel is assumed to vanish sufficiently fast as $s\rightarrow \infty$
\ba
D^{-1}(x,y)=\int_0^{+\infty} ds K(s;x,y;D)
\ea
The corresponding one loop effective action can be written as
\ba
W_{\rm{eff}}\equiv -\log Z^{\rm{1-loop}}=\frac{1}{2}\log \det D=-\frac{1}{2} \int_0^\infty \frac{ds}{s} K(s,D)
\ea
where $K(s,D)$ is defined as
\ba
 K(s,D)\equiv {\rm{Tr}} \exp(-sD)=\int d^n x \sqrt{g} K(s;x,x;D)
\ea

\subsection{Entanglement Entropy in BTZ Background}

In an elegant paper \cite{Mann:1996ze}, Mann and Solodukhin calculated the quantum correction to Bekenstein-Hawking entropy of the conical BTZ black hole using heat kernel method. The main result is reviewed in appendix. In this subsection we use their result to analyze the bulk entanglement entropy evolution.

Then the one loop effective action on $BTZ$ background can be calculated from heat kernel

\ba
W_{\rm{eff}}[BTZ] =-\frac{1}{2}\int_{\epsilon^2}^{+\infty}\frac{ds}{s} {\rm{Tr}}_{w=0} K_{BTZ}= W_{\rm{div}}[BTZ]+W_{\rm{fin}}[BTZ] \nonumber \\
\ea
The divergent part and the finite part of the effective action are given in appendix. The conical BTZ solution can be obtained from $AdS_3$ by identifications

 \ba
\theta \sim \theta+2\pi l \alpha,~~\psi \sim \psi+ \frac{2\pi r_+}{l}\sim  \psi+ \frac{2\pi l}{\beta_H}
 \ea
 For $\alpha \neq 1$, the solution after identifications which is named the conical BTZ has a conical singularity at $\rho=0$.
Using the heat kernels relation, the one loop effective action for conical BTZ has the form\cite{Mann:1996ze}
 \ba
W_{\rm{eff}}[cBTZ] =-\frac{1}{2}\int_{\epsilon^2}^{+\infty}\frac{ds}{s} {\rm{Tr}}_{w=0} K_{cBTZ}= W_{\rm{div}}[cBTZ]+W_{\rm{fin}}[cBTZ]
\ea
The finite part is
\ba \label{wfin}
W_{\rm{fin}}[cBTZ]=-\sum\limits_{n=1}^{+\infty} \frac{1}{4n} \frac{\sinh{\bar{A}_+n/\alpha}}{\sinh \bar{A}_+n} e^{-\sqrt{\mu} \bar{A}_+n}\sinh^{-2}\frac{\bar{A}_+ n}{2\alpha}
\ea
 The divergent part is given by (\ref{Wdivergent}). The UV divergent part of the effective action can be absorbed into the bare coupling constants in the classical action \cite{Susskind:1994sm}, as a result the bare coupling constants are replaced by the renormalized ones in the classical result. There are some recent discussions \cite{He:2014gva} in frame of string theory. The Ryu-Takayanagi entropy formula  becomes
 \ba
S^{cl}(A)=\frac{(\rm{Area)_{min}}}{4G^{\rm{ren}}_N}
\ea
Note that $G^{\rm{ren}}_N$ should be treated as macroscopically measurable constant.
The classical entanglement entropy, which is given by the minimal surface, is uninteresting to us since they remain the same under renormalization group flow from CFT$_A$ to CFT$_S$.
The bulk entanglement entropy, i.e. the one-loop correction of Bekenstein-Hawking entropy is given by
\ba
S_{\rm{bulk-ent}}=(\alpha \frac{\partial}{\partial \alpha}-1) W_{\rm{fin}}[cBTZ] \bigg|_{\alpha=1}
\ea
Using the finite part expression (\ref{wfin}), the bulk entanglement entropy of region $A_b$ whose corresponding boundary theory is CFT$_S$ can be calculated as
\ba \label{s}
S_{\rm{bulk-ent}}(\sqrt{\mu})=\sum\limits_{n=1}^{+\infty} \frac{1}{4n} \frac{e^{-\sqrt{\mu}\bar{A}_+n}}{\cosh{\bar{A}_+n} -1} \left[ 1+\bar{A}_+n\coth\bar{A}_+n - \frac{\bar{A}_+n \sinh\bar{A}_+n}{\cosh\bar{A}_+n-1} \right]
\ea
First consider the large $\bar{A}_+$ limit which corresponds to the big black hole limit. In this limit, the bulk entanglement entropy can be approximated in the following form
\ba
S_{\rm{bulk-ent}}(\sqrt{\mu}) &\simeq& -\ln \left(   1-   e^{-(\sqrt{\mu}+1)\bar{A}_+}  \right) \nonumber  \\
 &\simeq&  e^{-(\sqrt{\mu}+1)\bar{A}_+}
\ea
To obtain the bulk entanglement entropy due to boundary CFT$_A$, we analytically continue $\mu \rightarrow e^{2\pi i} \mu$. The bulk entanglement entropy due to boundary CFT$_A$ is
\ba
S_{\rm{bulk-ent}}(-\sqrt{\mu}) \simeq  e^{-(1-\sqrt{\mu})\bar{A}_+}
\ea
Under the double-trace deformation, the boundary theory flows from CFT$_A$ to CFT$_S$, the corresponding bulk entanglement entropy evolves from $S_{\rm{bulk-ent}}(-\sqrt{\mu})$ to $S_{\rm{bulk-ent}}(\sqrt{\mu})$. In the large black hole limit, the entropy difference is simply given by
\ba
\Delta S_{\rm{bulk-ent}} \simeq  e^{-(1-\sqrt{\mu})\bar{A}_+}- e^{-(\sqrt{\mu}+1)\bar{A}_+}
\ea
For the large black hole limit, i.e. $r_+\gg l$, we find that the bulk entanglement entropy decrease under the double-trace deformation which is consistent with the holographic $c$ theorem. The change of bulk entanglement entropy is positive which is consistent with pure AdS$_3$ case as well as $c$ theorem of two dimensional CFT.
Since the static BTZ black hole has positive specific heat and is thermodynamically stable for all values of $r_+$ \cite{Sarkar:2006tg}, the small $r_+$ limit also deserve attention. Note that the fact the BTZ black hole is thermodynamically stable for all values of $r_+$ is the necessary condition to use the off-shell approach to compute the entanglement entropy since in this approach the state of the black hole with arbitrary period of Euclidean time has a small deviation from the equilibrium.

For small $\bar{A}_+$ limit, which corresponds to the small black hole limit, we can approximate the sum in (\ref{s}) by integral
\ba \label{s}
S_{\rm{bulk-ent}}(\sqrt{\mu})\simeq \int^{+\infty}_{\bar{A}_+} dx \frac{1}{4x} \frac{e^{-\sqrt{\mu}x}}{\cosh x -1} \left[ 1+x\coth x - \frac{x \sinh x}{\cosh x-1} \right]
\ea
The integral can be expanded in $\bar{A}_+$ for small black hole limit
\ba \label{s}
S_{\rm{bulk-ent}}(\sqrt{\mu})\simeq \frac{\sqrt\mu}{12}-\frac{\ln\bar{A}_+}{12}+\mathcal{O}\left( \bar {A}_+^2\right)
\ea
The entropy difference is
\ba
\Delta S_{\rm{bulk-ent}} \simeq - \frac{\sqrt\mu}{12}-\frac{\sqrt\mu}{12}
\ea
which is negative different from the result of the large $\bar{A}_+$ limit.For small black hole limit, i.e. $r_+\ll l$, we find that the bulk entanglement entropy difference is negative, which means the bulk entanglement entropy is not monotonically decreasing under the renormalization flow.

\section{Conclusions}
In this paper, we study the bulk entanglement entropy in conical BTZ black bole background using the heat
kernel method. We also derive the entropy difference under boundary renormalization group
flow. The result depends on radius of black hole. For the large black hole limit, i.e. $r_+\gg l$, we find that the bulk entanglement entropy decrease under the double-trace deformation which is consistent with the holographic $c$ theorem. The change of bulk entanglement entropy is positive which is consistent with pure AdS$_3$ case as well as $c$ theorem of two dimensional CFT.
Since the static BTZ black hole has positive specific heat and is thermodynamically stable for all values of $r_+$ \cite{Sarkar:2006tg}, the small $r_+$ limit also deserve attention. Note that the fact the BTZ black hole is thermodynamically stable for all values of $r_+$ is the necessary condition to use the off-shell approach to compute the entanglement entropy since in this approach the state of the black hole with arbitrary period of Euclidean time has a small deviation from the equilibrium. For small black hole limit, i.e. $r_+\ll l$, we find that the bulk entanglement entropy difference is negative, which means the bulk entanglement entropy is not monotonically decreasing under the renormalization flow.

\section*{Acknowledgments}
We would like to thank for helpful discussions with Guo-Zhu Ning. We also thank the Email correspondence of Noburo Shiba for explaining their work. This work was supported by Natural Science Foundation of Hebei Province Grant No. A2014201169 and partly supported by the open project of State
Key Laboratory of Theoretical Physics with Grant No. Y3KF311CJ1.

\appendix
\renewcommand{\appendixname}{Appendix~\Alph{section}}

\section{Quantum Correction to BTZ Black Hole Entropy}

The quantum correction to Bekenstein-Hawking entropy of the conical BTZ black hole was calculated in an elegant paper \cite{Mann:1996ze}. In this appendix we review the main results that will be useful to our analysis. We start with $AdS_3$ in coordinates $(\sigma, \lambda, \phi)$
\ba \label{ads3s}
ds^2=d\sigma^2+l^2\sinh^2(\frac{\sigma}{l})(d\lambda^2+\sin^2\lambda d\phi^2)
\ea
which is a hyperbolic version of metric on the three-sphere
\ba
ds^2=d\sigma^2+l^2\sin^2(\frac{\sigma}{l})(d\lambda^2+\sin^2\lambda d\phi^2)
\ea
The point is that in coordinates $(\sigma,\lambda, \phi)$ of $AdS_3$, the geodesic distance of between two points with equal values of $\lambda$ and $\phi$ is given by
$\Delta \sigma= 	\left| \sigma-\sigma' \right| $ which is related to global $AdS$ coordinate (\ref {ads3}) via
\ba
\cosh \frac{\Delta\sigma}{l}=\cosh^2\rho\cosh\Delta \psi-\sinh^2\rho\cos \Delta\theta
\ea
In $(\sigma, \lambda, \phi)$ coordinates, the heat kernel equation in $AdS_3$ can be solved

\ba
K_{AdS_3}=\frac{1}{(4\pi s)^{\frac{3}{2}}}\frac{\sigma/l}{\sinh\frac{\sigma}{l}}\exp(-\frac{\sigma^2}{4s}-\mu \frac{s}{l^2})
\ea
where $\mu \equiv 1+m^2l^2$.

Since BTZ solution can be constructed from $AdS_3$ by combination of identifications (\ref{ads3}), the heat kernel $K_{BTZ}$ on BTZ black hole background can be constructed from $K_{AdS_3}$ on $AdS_3$ background as infinite sum over images
\ba
 K_{BTZ}=\sum\limits_{n=-\infty}^{+\infty}K_{AdS_3}(\sigma_n,s), \nonumber
 \ea
 \ba
 \cosh \frac{\Delta\sigma}{l}=\cosh^2\rho\cosh\Delta \psi_n-\sinh^2\rho\cos \Delta\theta,  \nonumber
 \ea
 \ba
 \Delta \psi_n=\psi-\psi'+2\pi\frac{r_+}{l}n,~ \Delta\theta=\theta-\theta',
\ea
For further application, define the integral of $K_{BTZ}$
\ba
{\rm{Tr}}_w K_{BTZ}\equiv \int_{BTZ} d\mu_x K_{BTZ}(\rho=\rho',\psi=\psi',\theta=\theta'+w)\nonumber
\ea
\ba\label{traceK}
d\mu_x=l^3\cosh\rho\sinh\rho d\rho d\theta d\psi
\ea
The volume on $BTZ$ can be written as
\ba
V_{BTZ}=\int_{BTZ} d\mu_x=l^3\int_0^{2\pi} d\theta \int_0^{\frac{2\pi r_+}{l}}d\psi \int_0^{+\infty} d\rho\cosh\rho\sinh\rho
\ea
Introduce the cutoff surface at large value of radius $\rho_c$, then the volume on $A_b$ becomes
\ba
V_{BTZ}=4\pi^2 l^2r_+\int_0^{\rho_c}d\rho \cosh\rho\sinh\rho\equiv V_r/\beta
\ea
After integration, E.q.(\ref{traceK}) reads
\ba
{\rm{Tr}}_w K_{BTZ}= \frac{e^{-\mu \bar{s}}}{(4\pi\bar{s})^\frac{3}{2}}\left [V_w+2\pi\bar{s}(\frac{2\pi r_+}{l})\sum\limits_{n=1}^{+\infty} \frac{\exp (\frac{-\Delta \psi_n^2}{4\bar{s}})}{\sinh^2\frac{\Delta \psi_n}{2}+\sin^2\frac{w}{2} }  \right] \nonumber
\ea
\begin{eqnarray}V_w=
\begin{cases}
\frac{V_{BTZ}}{l^3}, &w=0  \cr2\pi \left( \frac{2\pi r_+}{l} \right) \frac{\bar{s}}{2} \frac{1}{\sin^{2}\frac{w}{2}}, &w \neq 0\end{cases}
\end{eqnarray}
where we have defined $\bar{s}=s/l^2$ and $\Delta \psi_n=n2\pi r_+/l =n2\pi l/\beta$.

Then the one loop effective action on $BTZ$ can be calculated from heat kernel

\ba
W_{\rm{eff}}[BTZ] =-\frac{1}{2}\int_{\epsilon^2}^{+\infty}\frac{ds}{s} {\rm{Tr}}_{w=0} K_{BTZ}= W_{\rm{div}}[BTZ]+W_{\rm{fin}}[BTZ] \nonumber \\
\ea
where $\bar{A}_+ = 2\pi r_+/l=2\pi l/\beta$, and the divergent part of the effective action takes the form
\ba
 W_{\rm{div}}[BTZ]&=& -\frac{1}{2}\frac{1}{(4\pi)^\frac{3}{2}} V_{BTZ}\int_{\epsilon^2}^{+\infty} ds s^{-\frac{5}{2}} e^{-\mu \bar{s}} \nonumber \\
 &=& -\frac{1}{(4\pi)^\frac{3}{2}} V_{BTZ} \left[ \frac{1}{3\epsilon^3}-\frac{2 \bar{\mu}}{3\epsilon} +\frac{2}{3} \bar{\mu}^{\frac{3}{2}}\sqrt{\pi}+ \mathcal{O}(\epsilon) \right]
\ea
where we have defined $\bar{\mu}=\mu/l^2$. The finite part of the effective action is
\ba
W_{\rm{fin}}[BTZ]=-\sum\limits_{n=1}^{+\infty} \frac{1}{4n} e^{-\sqrt{\mu} \bar{A}_+n}\sinh^{-2}\frac{\bar{A}_+ n}{2}
\ea

The conical BTZ solution can be obtained from $AdS_3$ by identifications

 \ba
\theta \sim \theta+2\pi l \alpha,~~\psi \sim \psi+ \frac{2\pi r_+}{l}\sim  \psi+ \frac{2\pi l}{\beta_H}
 \ea
 For $\alpha \neq 1$, the solution after identifications which is named the conical BTZ has a conical singularity at $\rho=0$.
The heat kernel in conical BTZ can be expressed in terms of the heat kernel in BTZ \cite{Dowker:1977zj} \cite{Mann:1996ze}
 \ba
 K_{cBTZ}(x,y,s)=K_{BTZ}(x,y,s)+\frac{1}{4\pi\alpha} \int_{\Gamma}\cot{\frac{w}{2\alpha}}\times K_{BTZ}(\theta-\theta'+w,s)dw
 \ea
 where the contour $\Gamma$ consists of two vertical lines, going from $-\pi+i\infty$ to $-\pi-i\infty$ and from $\pi-i\infty$ to $\pi+i\infty$. Using the heat kernels relation, the one loop effective action for conical BTZ has the form\cite{Mann:1996ze}
 \ba
W_{\rm{eff}}[cBTZ] =-\frac{1}{2}\int_{\epsilon^2}^{+\infty}\frac{ds}{s} {\rm{Tr}}_{w=0} K_{cBTZ}= W_{\rm{div}}[cBTZ]+W_{\rm{fin}}[cBTZ]
\ea
The finite part is
\ba
W_{\rm{fin}}[cBTZ]=-\sum\limits_{n=1}^{+\infty} \frac{1}{4n} \frac{\sinh{\bar{A}_+n/\alpha}}{\sinh \bar{A}_+n} e^{-\sqrt{\mu} \bar{A}_+n}\sinh^{-2}\frac{\bar{A}_+ n}{2\alpha}
\ea
 The divergent part is
\ba \label{Wdivergent}
 W_{\rm{div}}[cBTZ]= -\frac{1}{2}\frac{1}{(4\pi)^\frac{3}{2}} \left[ V^c_{A_b}\int_{\epsilon^2}^{+\infty} ds s^{-\frac{5}{2}} e^{-\mu \bar{s}}+ \frac{1}{2}A_+ (2\pi\alpha) c_2(\alpha) \int_{\epsilon^2}^{+\infty} ds s^{-\frac{3}{2}} e^{-\mu \bar{s}} \right] \nonumber \\
 = -\frac{1}{(4\pi)^\frac{3}{2}} \left[ V_{A_b} \left( \frac{1}{3\epsilon^3}-\frac{2 \bar{\mu}}{3\epsilon} +\frac{2}{3} \bar{\mu}^{\frac{3}{2}}\sqrt{\pi}+ \mathcal{O}(\epsilon)  \right) + \frac{1}{2}A_+ (2\pi\alpha) c_2(\alpha)\left(  \frac{1}{\epsilon} -\frac{\sqrt{\pi \mu}}{l} + \mathcal{O}(\epsilon)       \right) \right]\nonumber\\
\ea

\end{document}